\documentclass[twocolumn,superscriptaddress,showpacs,prd,aps,amsmath,amssymb,nofootinbib]{aastex631}
\usepackage{natbib}
\setcitestyle{square,numbers}
\usepackage{graphicx,color}
\usepackage{amsmath,amssymb}
\usepackage{verbatim}
\usepackage{hyperref}
\usepackage{float}
\usepackage{wasysym}
\usepackage{amssymb,graphicx}
\usepackage{epsfig}
\usepackage{psfrag}
\usepackage{dsfont}
\usepackage{amsfonts}
\usepackage{mathrsfs}
\usepackage{multirow}
\usepackage{times}
\usepackage{soul}
\usepackage{orcidlink}

\begin{document}

\title{The Eras Tour: Mapping the Eras of Taylor Swift to the Cosmological Eras of the Universe}

\author{Jane C. Bright \orcidlink{0000-0003-2042-126X}}
\affiliation{Department of Physics, Grinnell College, Grinnell, IA}


\begin{abstract}

This paper explores an unexpected yet compelling parallel between the evolution of the universe, as described by cosmological eras, and the artistic evolution of Taylor Swift, delineated by her distinct album eras. By mapping key characteristics and transitions in the universe's history to corresponding themes and milestones in Swift's career, I offer a novel perspective on both. I culminate with predictions for Swift's future work and dare to ask a question of cosmic importance: Could Taylor Swift's thirteenth album hold the secret to the universe's ultimate destiny?

\end{abstract}


\section{Introduction}

We currently understand the history of our universe through the Big Bang Theory, which follows our observations that the universe is expanding, and proposes distinct eras of evolution that reach back all the way to the first instant of the birth of our universe. These eras mark distinct characteristics of what the universe was like at these different times, and mark important transitions and transformations in the universe's structure, composition, and behavior. 

In the common vernacular, the term ``era" has experienced a rise in popularity due to the popular musical artist, Taylor Swift, and her world tour entitled ``The Eras Tour" which took place from 2023 to 2024 \cite{erastour}. At the time of writing, Taylor Swift has released eleven studio albums, with her debut album released in 2006, and her most recent album \textit{The Tortured Poets Department}, released in April of 2024. The fan base of Taylor Swift, colloquially referred to as ``Swifties", has noted the distinct theme, musical style, visual style, and even associated color, for each of Swift's albums, and have thus come to refer to each album not just as a list of songs, but as a clearly distinguishable era. 

For each of Swift's albums beginning with her second studio album \textit{Fearless} released in 2008, through her sixth studio album \textit{reputation} released in 2018, Swift put out an accompanying tour following the release of the album that showcased each era. Due to the global shutdown from COVID-19, Swift was unable to follow this pattern of touring each individual album, and released four new albums (\textit{Lover}, \textit{folklore}, \textit{evermore}, and \textit{Midnights}) before she was able to go on tour again. This led to Swift's creation of ``The Eras Tour" which, rather than highlighting a single album, would bring together her entire body of work through meticulously crafted vignettes of each of her eras, including many of the most popular songs from each album, set changes, costume changes, and distinct shifts of aesthetic through each part of the 3.5 hour show. In the second year of the tour, after the release of \textit{The Tortured Poets Department}, the set list and sequencing of the show was rearranged to allow for the inclusion of the new era. 

The tour was the biggest in world history, and the discussion of which of Swift's eras a person was experiencing in their own lives became common in popular culture \cite{erasrecords}. The use of the term ``era" was then propelled from just Swift's eras to the discussion of personal eras to generally encompass phases of one's life, even leading to a new definition of the word era in the Collins dictionary as ``a period of one's life or career considered as being of a distinctive character" \cite{eradef}. 

As the term era had been used to describe the cosmological eras of the evolution of the universe prior to the term's rise in popular use due to Taylor Swift, it seems only natural to form a link between the two uses. In this paper, I will map the eras of Taylor Swift to the cosmological eras of the universe. In section \ref{sec:cosmological} I will provide a brief outline of the cosmological eras, their timelines, and their distinct characteristics. In section \ref{sec:swift} I will provide a brief outline of the Taylor Swift eras, their major themes, and the importance they played in Swift's career. In section \ref{sec:map} I will introduce my mapping between the cosmological and Taylor Swift eras and my reasoning for the mapping. In section \ref{sec:TS12} I will use our understanding of the cosmological eras to make predictions for Taylor Swift's next era and album, which will be her twelfth studio album and referred to by Swifties as \textit{\textit{TS12}} (short for Taylor Swift 12) until an official name is announced. Finally, in section \ref{sec:TS13} I will describe how Swift's thirteenth studio album (again colloquially referred to as \textit{TS13}) will predict the ultimate fate of the universe.  

\section{The Cosmological Eras}\label{sec:cosmological}

\textit{\textbf{Planck Era}}: This is the earliest known era, occurring from the Big Bang to approximately $10^{-43}$ seconds after.
During this time, the four fundamental forces of nature (gravity, strong nuclear force, weak nuclear force, and electromagnetism) are believed to have been unified into a single force, making it a time of ultimate simplicity in the universe. Our current understanding of physics is insufficient to fully describe this era.

\textit{\textbf{Grand Unification Theory (GUT) Era}}: Lasting from $10^{-43}$ to $10^{-36}$ seconds after the Big Bang. The GUT era began when the temperature of the universe dropped to a low enough temperature (approximately $10^{32}$ K)  that gravity was able to break away and become distinct from the other forces, which remained unified as the GUT force. 

\textit{\textbf{Inflation}}: The GUT era ended when the strong force separated out, splitting the GUT force into the strong force and the electroweak force, releasing an enormous amount of energy. This energy caused a short and dramatic expansion of the universe. 

\textit{\textbf{Electroweak Era}}:  Spanning $10^{-36}$ to $10^{-12}$ seconds after the Big Bang, the electroweak era was the time where the electromagnetic and weak forces were still merged. The universe was filled with intense radiation during this time that spontaneously produced particles and antiparticles that would almost immediately annihilate and turn back into photons. This era ended when the universe cooled to the point that the electromagnetic and weak forces separated, and our universe has had four distinct forces ever since. 

\textit{\textbf{Particle Era}}: This era is marked by the time after the  electromagnetic and weak forces split, but the universe was still hot enough for the spontaneous production of particles, which lasted until $10^{-3}$ seconds after the Big Bang. Photons produced many exotic particles, such as quarks, which we do not find free in the universe today. By the end of this era, all quarks were bound into protons and neutrons. We conclude that there must have been a slight imbalance between the numbers of particles and anti-particles produced given that at the end of the particle era the universe was left with some matter and not all matter annihilated with anti-matter to leave only photons in the universe. 

\textit{\textbf{Era of Nucleosynthesis}}: This era occurred from $10^{-3}$ seconds to 3 minutes after the Big Bang. The universe was hot enough for nuclear fusion to occur, and protons and neutrons combined to form the first atomic nuclei, primarily hydrogen and helium. At the end of this era, the chemical makeup of the universe was about 75\% hydrogen, 25\% helium, and trace amounts of lithium and deuterium. 

\textit{\textbf{Era of Nuclei}}: This era occurred from 3 minutes to roughly 380,000 years after the Big Bang.
The universe was a hot dense plasma of nuclei, electrons and photons.
The universe was still too hot for electrons and nuclei to bind to form neutral atoms. Photons bounced between unbound electrons, never able to travel very far between collisions, making the universe opaque.

\textit{\textbf{Era of Atoms}}: 
The universe cooled enough for electrons to combine with protons and other atomic nuclei.
This resulted in the formation of neutral atoms, primarily hydrogen and helium.
Before this era, the universe was a hot, dense plasma, opaque to light.
The formation of neutral atoms made the universe transparent, allowing photons to travel freely.
The photons released during recombination form the cosmic microwave background (CMB) radiation, the ``afterglow" of the Big Bang.
This is the oldest light we can detect.

\textit{\textbf{Cosmic Dark Ages}}: 
Following the formation of neutral atoms, the universe was filled with mostly hydrogen and helium. No stars or other light sources existed yet, making the universe dark and relatively featureless.
This period was a time of gradual evolution, where gravity began to slowly clump matter together, eventually leading to the formation of the first stars.

\textit{\textbf{Cosmic Dawn}}: 
This is the era when the very first stars ignited in the universe, ending the preceding Cosmic Dark Ages. It marks the point where the universe transitioned from a dark, relatively simple state to one filled with light and the beginnings of complex structures. This period saw the initial formation of stars from clumps of hydrogen and helium, which began to illuminate the cosmos and initiate the formation of galaxies. The light from these earliest stars is highly redshifted due to the universe's expansion, providing astronomers with a way to study this pivotal time.

\textit{\textbf{Reionization}}: 
The first stars and quasars emitted intense ultraviolet radiation.
This radiation reionized the neutral hydrogen that filled the universe, stripping electrons from the atoms.
The universe transitioned from a predominantly neutral state to an ionized state.
This reionization process made the universe even more transparent to ultraviolet light.

\textit{\textbf{Era of Galaxies}}: 
Gravity caused the clumping of matter into increasingly larger structures.
These structures eventually formed galaxies, vast collections of stars, gas, dust, and dark matter.
Within galaxies, stars continued to form from collapsing clouds of gas and dust.
This process continues to this day.
Galaxies have evolved and changed over time through mergers, interactions, and star formation.
Different types of galaxies, such as spiral, elliptical, and irregular, emerged.
The universe continues to expand during this era, and galaxies are moving further apart.
This era continues to this day, and is the era that we currently live in.

\section{The Taylor Swift Eras}\label{sec:swift}

Taylor Swift's musical career can be broken down into her eleven studio albums, each distinguished by its own distinct era:

\textit{\textbf{Taylor Swift}}: Swift's debut, self-titled album was released in 2006 when Swift was 16-years-old. It is her most country genre album, and features songs with themes of teenage love and loss. Visually, it is associated with a light green color, cowboy boots, and other county style aesthetics. 

\textit{\textbf{Fearless}}: While sticking to her country roots, Fearless had a more pop-country sound than her debut album. Thematically, the album has a fairytale characteristic, highlighted by songs such as ``Love Story" and ``White Horse" which won Grammys for Best Country Song and Best Female Country Vocal Performance \cite{grammy}. \textit{Fearless} also won Grammys for Best Country Album and Album of the Year \cite{grammy}. While her debut album had been popular, Fearless was truly her breakaway album and launch into stardom. The color gold is associated with this album.

\textit{\textbf{Speak Now}}: Continuing her pop-country sound, \textit{Speak Now} truly highlighted Swift's songwriting and lyrical prowess as she wrote the album entirely herself. The writing style is very personal, confessional, and pointed, showcasing her ability to write, create, and tell stories. The visual style of the era follows from the album cover, which depicts her in a whimsical purple dress. 

\textit{\textbf{Red}}: This album showcased further movement towards the pop side of her pop-country sound. The era depicts emotional highs and lows, exploring passionate and sometimes tumultuous relationships, and reflects the maturity and depth of her moving into adulthood in songs like ``All Too Well" and ``State of Grace" as well as the frivolity and freedom seeking of youth in songs such as ``22" and ``We Are Never Ever Getting Back Together." The era came to be associated with autumn, scarves, and of course, the color red.  

\textit{\textbf{1989}}: This album solidified her full transition into the pop genre as demonstrated by the Grammy win for Best Pop Vocal Album as well as Album of the Year \cite{grammy}. It is characterized by synth-pop sounds and anthems of independence and city life, and is associated with light blue, New York City, and seagulls (as featured on the album cover). This era saw her rise from a country singer to a true pop star. 

\textit{\textbf{reputation}}: This era took a darker, edgier turn, with themes of fame, revenge, and reclaiming her narrative. Following a period of intense media attention and public feuds, Taylor retreated from the public eye.
\textit{Reputation} was her response to this scrutiny. She did very little press during this time, with the motto ``there will be no explanation, there will just be reputation." Visually, this album is associated with dark tones, particularly black, and the symbol of the snake. 

\textit{\textbf{Lover}}: 
The Lover era represented a 180 from the previous darkness of reputation, leaning heavily into light pastel colors, sunrise skies, and a generally warm, bright aesthetic. The era's themes were much more optimistic, upbeat, and warm. 

\textit{\textbf{folklore}}: 
The folklore era took place in 2020 in the wake of the global COVID-19 pandemic. The album echoed the experiences of departing from our regular lives and being in isolation. The album featured an indie folk and alternative sound, with a focus on storytelling and character-driven narratives. The album's production was largely acoustic and introspective, creating a more intimate and atmospheric listening experience. This era was one of Swift's first big departures from autobiographical storytelling and a focus on fictional stories and ``folklore." This album was a surprise album, announced just hours before its release. 

\textit{\textbf{evermore}}: 
Released just months after \textit{folklore}, \textit{evermore} was conceived as a continuation of that album's creative exploration.
It maintained the narrative-driven songwriting and fictional storytelling of \textit{folklore} and was considered to be a ``sister album" to \textit{folklore}. It felt like a creative extension, and continuation of the artistic freedom that was expressed in \textit{folklore}. This was a departure from Swift's usual trend of making a clean break in style and aesthetic between each album, and was born from Swift's desire to just keep writing and telling stories during the pandemic time. 

\textit{\textbf{Midnights}}: Midnights was a concept album centered around the theme of ``what keeps you up at night" and each song represents a different ``midnight" thought from some time in Swift's life. While the sound is more pop-driven with an electronic and synth-pop sound, the lyrics retain the introspective and narrative-driven storytelling of her recent work, delving into personal experiences, anxieties, and reflections.

\textit{\textbf{The Tortured Poets Department}}: 
The album is marked by its intense emotional vulnerability, with Taylor laying bare her feelings and experiences.
It is very raw, and honest.
The lyrics are highly poetic and literary, filled with metaphors and symbolism.
The name of the album sets this tone.
A central theme is the exploration of heartbreak, with songs delving into the pain, confusion, and aftermath of broken relationships. The album was released as a double album, with ``The Anthology" adding many additional songs.

\section{Mapping the Eras}\label{sec:map}

\begin{figure*}
    \centering
    \includegraphics[width=\textwidth]{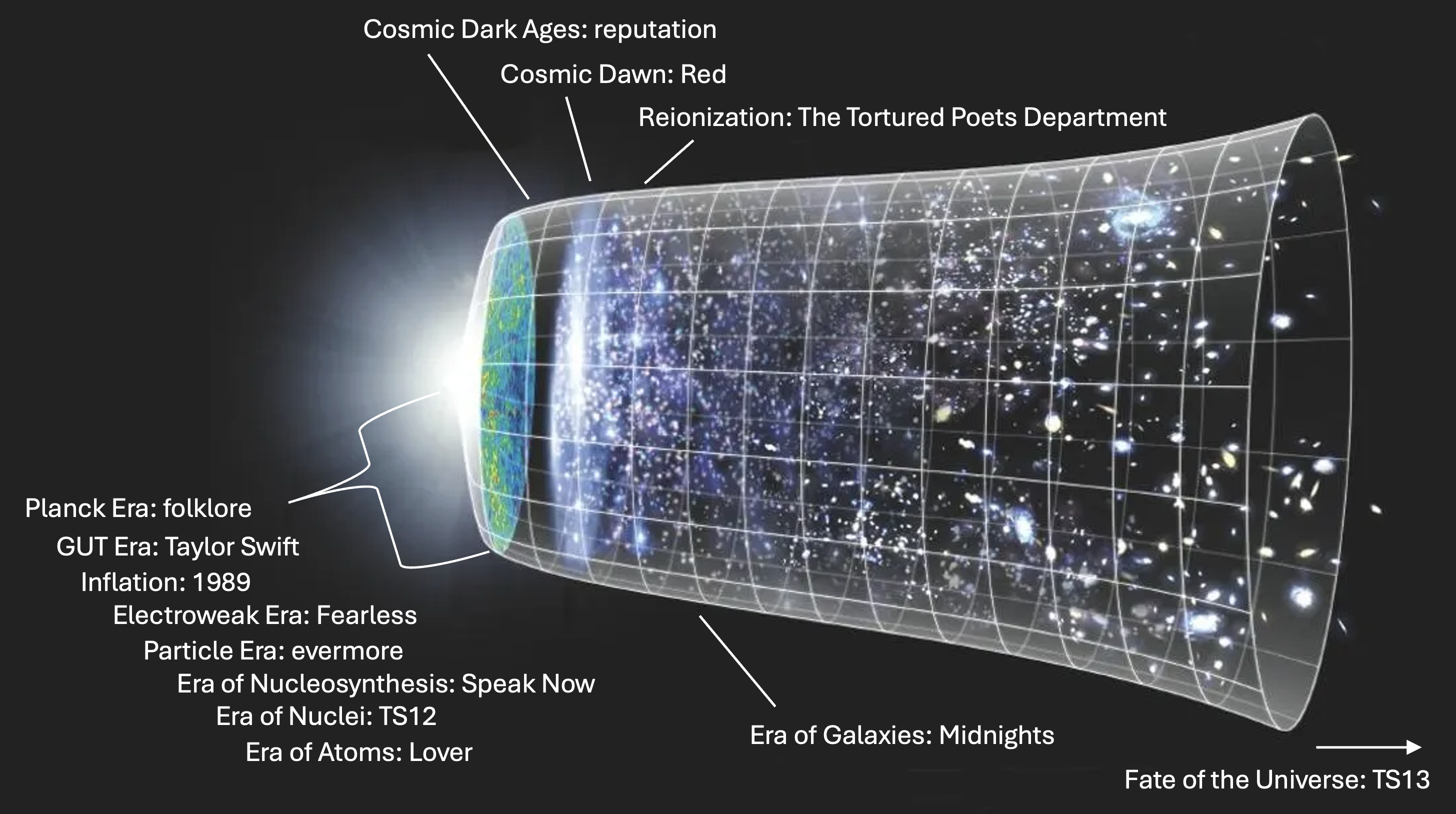}
    \caption{Artist's impression of the timeline of the universe, summarizing the mapping between the cosmological eras the Taylor Swift eras. Image adapted from NASA/WMAP.}
    \label{fig:cosmology}
\end{figure*}

I will now create a mapping between the cosmological eras and the Taylor Swift Eras. As seen in the performance of ``The Eras Tour" and the order of her re-release of her first six albums, Swift does not always present her work in the chronological order of its original release, but in the order to create a compelling experience and flow. Likewise, I will not simply create a chronological mapping of cosmological and Taylor Swift eras, but attempt to create the mapping that links the themes and distinctions of each era. This mapping is summarized in Fig. \ref{fig:cosmology}.

\textit{\textbf{Taylor Swift (Debut): GUT Era:}}
The GUT era marks when gravity first separates from the other fundamental forces, perfectly mirroring Taylor's debut. It was her first step out on her own, a moment of distinct individuality and the beginning of her own gravitational pull on the music world. Just as the GUT Era represents the emergence of a distinct force, \textit{Debut} marked Taylor's emergence as a distinct artist.

\textit{\textbf{Fearless: Electroweak Era:}}
The electroweak era is characterized by the separation of the electroweak force into the weak nuclear force and electromagnetism. It's a period of differentiation, where previously unified forces become distinct. There is still a large amount of energy, and rapid change.
 \textit{Fearless} marks a significant step in Taylor's artistic evolution as she began to experiment with her sound, expanding beyond her country roots while still maintaining her core songwriting style. This is a clear differentiation of her artistic forces.
 The album showcases a growing complexity in her songwriting and production. While still rooted in country, it hints at the pop sensibilities that would later dominate her work. This mirrors the increasing complexity of the universe as forces separate, but have not completely differentiated yet.
\textit{Fearless} is an album filled with youthful energy and a sense of transition. It captures the feeling of moving from one phase to another, mirroring the dynamic changes of the Electroweak Era.
The rapid growth of her fan base, and the rapid increase in her popularity during this era, mirrors the rapid changes of the electroweak era.

\textit{\textbf{Speak Now: Era of Nucleosynthesis:}}
In the era of nucleosynthesis, protons and neutrons combined to form atomic nuclei, building something entirely new and creating the building blocks of the universe. \textit{Speak Now}, entirely self-written, represents Swift building something entirely new from scratch. She was building her own artistic foundation, crafting each song independently. She was building her reputation as a song writer. This era is marked by the creation of new, more complex structures from simpler ones, which aligns with Taylor constructing a cohesive and personal album from her own creative ``particles."

\textit{\textbf{Red: Cosmic Dawn:}}
Cosmic Dawn was a period of intense energy release, as the first stars ignited.
\textit{Red} is known for its raw, passionate emotions, its soaring highs and devastating lows. This emotional intensity mirrors the energetic birth of the first stars. Cosmic dawn output the very first starlight in the universe, perfectly aligned with Swift's song ``Starlight" off of \textit{Red}. The song ``Treacherous" also has the lyrics ``'til the gravity's too much" which is exactly the hallmark of the gravitational collapse of the first luminous object in the universe due to gravity. These first luminous objects are also the most highly redshifted objects in the universe, perfect for the album \textit{Red}. 

\textit{\textbf{1989: Inflation:}}
Cosmic inflation was a period of extremely rapid expansion in the early universe, where the universe's size increased exponentially in a very short time.
\textit{1989} represents Taylor's rapid rise to pop superstardom, a period of explosive growth in her popularity and influence.
Just as inflation saw a dramatic break away from the previous state of the universe, \textit{1989} marked Taylor's decisive break away from her country roots and her embrace of pop music.
The strong force breaking away from the electroweak force, created a massive change, and the change from country artist to pop artist created a massive change in the music world.
The sheer scale of the expansion in both cases is a striking parallel.

\textit{\textbf{reputation: Cosmic Dark Ages:}}
During the Cosmic Dark Ages, the universe was filled with neutral hydrogen and was largely devoid of light.
\textit{reputation} similarly felt like a period of ``darkness" for Taylor, marked by media scrutiny, public feuds, and a deliberate withdrawal from the public eye.
Just as the Dark Ages were a period of relative quiet and unseen processes, \textit{reputation} was a time of internal processing and rebuilding for Taylor, away from the bright lights of her previous persona.
Also, the dark visuals, and the tone of the album, matches the dark ages.

\textit{\textbf{Lover: Era of Atoms:}}
In the era of atoms, recombination made the universe transparent to light, allowing photons to travel freely. This mirrors the ``daylight" theme (and song) in \textit{Lover}, symbolizing clarity, openness, and a newfound sense of visibility.
The overall warm, bright, and airy feel of the album reflects this newfound transparency. 
Recombination was the first time electrons and nuclei bound together to form neutral atoms. This parallels the themes of connection, partnership, and commitment in \textit{Lover}.
The CMB is the afterglow of the Big Bang, a remnant of the universe's early heat.
The song ``Afterglow" captures the feeling of lingering warmth and the aftereffects of an intense emotional experience, mirroring the CMB's afterglow.

\textit{\textbf{folklore: Planck Era:}} 
The Planck Era is the earliest known period, where our understanding of physics breaks down. It's a time of ultimate simplicity, where all forces are unified, and the universe exists in a state we can barely comprehend.
\textit{folklore} is characterized by its stripped-down production, its focus on storytelling, and its exploration of mysterious and timeless themes.
Just as the Planck Era is shrouded in mystery, \textit{folklore} evokes a sense of ancient tales and hidden meanings.
The idea of ultimate simplicity can be found in the more acoustic nature of the album, and the focus on the bare bones of the stories being told.
The unknown nature of the Planck Era, and the fictional stories of \textit{folklore}, connect well. \textit{folklore}'s release as a surprise album also echoes the Planck Era as the first in the universe after the Big Bang, occurring very suddenly.

\textit{\textbf{evermore: Particle Era:}}
The Particle Era was a time of intense energy, with particles constantly being created and annihilated. It was a period of rapid change and dynamic activity.
\textit{evermore}, as a sister album to \textit{folklore}, feels like an extension of that creative energy. Swift continued to write stories, explore characters, and delve into the depths of her imagination.
The sheer volume of stories and the continuous flow of creative output in \textit{evermore} can be seen as analogous to the constant creation and annihilation of particles.
The particle era is rapid and very energetic, and the rapid release of evermore, right after folklore, shows that rapid energy.
The flow of energy, and the constant state of change, is a good way to describe both.

\textit{\textbf{Midnights: Era of Galaxies:}}
The Era of Galaxies is characterized by the formation of complex structures, the accumulation of vast amounts of matter, and the ongoing evolution of the universe.
\textit{Midnights} feels like a collection of intricate, interconnected thoughts and reflections that have built up and grown over the course of Swift's life, not just distilling a single moment in time. This echoes the vast span of the era of galaxies. 
Galaxies are vast collections of stars, gas, and dust held together by gravity. \textit{Midnights} is a collection of songs, held together by the theme of midnight thoughts. The imagery of \textit{Midnights} including dark blues and stars is of course reminiscent of the of images of galaxies, and the song ``Bigger Than the Whole Sky" feels very related to the expanse of a galaxy. 

\textit{\textbf{The Tortured Poets Department: Reionization:}}
Reionization is the period after Cosmic Dawn, where the radiation from the first stars and quasars reionized the neutral hydrogen in the universe, essentially breaking apart and transforming the remaining neutral atoms.
\textit{The Tortured Poets Department} feels like a period of intense emotional and creative ``reionization" for Swift. The relationships and bonds she thought would be permanent have been broken apart, and she's dissecting, breaking apart, and re-evaluating past experiences, relationships, and even her own artistic persona.
The album is full of raw, unfiltered emotions, and a sense of breaking down previous walls or perceptions. This ``breaking apart" perfectly mirrors the reionization of hydrogen.

\section{\textit{TS12} Predictions}\label{sec:TS12}

Following the patterns presented above, it follows that Swift's next album must map to the Era of Nuclei, as this era does not have a pairing within her present body of work. We may then use our understanding of the universe to make predictions for what we should expect from Swift's twelfth studio album, referred to by Swifties as \textit{\textit{TS12}} until a title is announced. The Era of Nuclei saw the universe as a hot, dense plasma. The universe was extremely energetic and charged as neutral atoms could not yet form.
It was an era of intense interactions with particles and photons constantly colliding and interacting.
We can thus predict that \textit{\textit{TS12}}  will be a very energetic album, perhaps with a more intense or experimental sound. Think heavily produced, dense layers of sound, and a lot of driving rhythms. The ``charged" atmosphere of the Era of Nuclei might translate to an album filled with heightened emotions, perhaps dealing with conflict or passion. The universe being opaque during the Era of Nuclei may also lead to songs with hidden or unclear meanings. As nuclei are the building blocks of the universe, \textit{\textit{TS12}} may represent laying the foundations for the next stage of her career or her personal life, as many fans expect to see her settle down with her boyfriend, Travis Kelce.

\section{\textit{TS13} and the Fate of the Universe}\label{sec:TS13}

It then follows naturally that Swift's thirteenth studio album, \textit{TS13}, would map to the final era of the universe and the ultimate fate of the universe. There are several proposed fates of the universe, including: \textit{ Big Freeze:}
The universe continues to expand, eventually reaching a state of maximum entropy.
Stars burn out, black holes evaporate, and the universe becomes cold and empty.
\textit{Big Rip:}
The expansion of the universe accelerates to the point where it tears apart galaxies, stars, planets, and eventually atoms themselves.
\textit{Big Crunch:}
The expansion of the universe reverses, and everything collapses back into a singularity.

When Swift releases \textit{TS13}, we may then make predictions for the fate of the universe based upon the themes and styles present in \textit{TS13} to give conclusive evidence for which fate matches to \textit{TS13}. It is only natural Swift's thirteenth album would hold such cosmic importance, as the number 13 is incredibly important to Swift personally and by extension to the Swiftie fanbase. 
Taylor Swift was born on December 13th. 
Taylor has publicly stated that 13 is her lucky number. 
She has noted that the number 13 frequently appears in her life, often in unexpected and positive contexts.
During her early career, she would often write the number 13 on her hand before performances.
Over time, 13 has become a symbol of connection between Taylor and her fans. Swifties recognize it as a shared symbol, a kind of inside joke. The number 13 represents a sense of fate, destiny, and the interconnectedness of events.
Therefore, if \textit{TS13} were to embody the final era of the universe, it could be seen as Taylor wielding a kind of cosmic influence.
\textit{TS13} could be a cosmic catalyst:
instead of simply reflecting the universe's fate, \textit{TS13} could determine it.
The album's themes, emotions, and sonic landscapes could act as a kind of cosmic trigger, setting in motion the final events of the universe.

\section{Conclusions}
In this paper, we have embarked on a cosmic journey guided by the light of Taylor Swift's discography, presenting a novel mapping between the cosmological eras of the universe and the distinct eras of her musical journey. By drawing parallels between the evolution of the cosmos and the artistic development of a pop icon, I have shown that both are characterized by distinct periods of change, energy, and transformation. Perhaps boldly (or perhaps foolishly), I have attempted to bridge the gap between pop culture and astrophysics. Furthermore, I have dared to gaze into the future, speculating that Taylor Swift's upcoming albums may hold cosmic significance, with \textit{TS13} potentially revealing the ultimate fate of the universe. While this analysis may seem unconventional, it highlights the pervasive influence of Taylor Swift in contemporary culture. And, let's be honest, if anyone is destined to hold the key to the universe's fate, it might just be her.

\section*{Acknowledgments}

I gratefully acknowledge the students in the Fall 2024 PHY-116 ``The Universe and its Structure" course at Grinnell College who inspired the idea for this paper. Of course, the greatest acknowledgment is owed to Taylor Alison Swift: thank you for the eras, the inspiration, and the cosmic mystery. The universe, it turns out, is a Swiftie.

\bibliographystyle{apalike}
\bibliography{Bibliography}

\end{document}